\documentclass[aps,preprint,amsmath,superscriptaddress,nofootinbib,byrevtex,showkeys,showpacs]{revtex4}
\usepackage{graphicx}
\DeclareGraphicsExtensions{.eps}
\usepackage{subfigure}

\begin{document}


\title{Reflections on Modern Macroeconomics: Can We Travel Along a Safer Road?}\thanks{The title echoes \cite{Kirman2006}, to which we are indebted for some of the observations here reported.}
\author{E. Gaffeo}
\affiliation{Department of Economics and CEEL, University of Trento, Via Inama 5, 38100 Trento, Italy}
\author{M. Catalano}
\author{F. Clementi}
\email[Corresponding author: ]{fabio.clementi@uniroma1.it}
\affiliation{Department of Economics, Universit\`a Politecnica delle Marche, Piazzale Martelli 8, 60121 Ancona, Italy}
\author{D. Delli Gatti}
\affiliation{Institute of Economic Theory and Quantitative Methods, Catholic University of Milan, Via Largo Gemelli 1, 20123 Milan, Italy}
\author{M. Gallegati}
\affiliation{Department of Economics, Universit\`a Politecnica delle Marche, Piazzale Martelli 8, 60121 Ancona, Italy}
\author{A. Russo}
\affiliation{Scuola Normale Superiore, Piazza dei Cavalieri 7, 56126 Pisa, Italy}
\date{\today}


\begin{abstract}
In this paper we sketch some reflections on the pitfalls and inconsistencies of the research program\textemdash currently dominant among the profession\textemdash aimed at providing microfoundations to macroeconomics along a Walrasian perspective. We argue that such a methodological approach constitutes an unsatisfactory answer to a well-posed research question, and that alternative promising routes have been long mapped out but only recently explored. In particular, we discuss a recent agent-based, truly non-Walrasian macroeconomic model, and we use it to envisage new challenges for future research.
\end{abstract}
\pacs{89.65.Gh}
\keywords{equilibrium; agent-based economics; emergence}
\maketitle


\section{Introduction}
In a recent paper \cite{Keen2003}, Steve Keen warns econophysicists against using modern economic theory as a foundation for their work, boosting them instead to learn directly from the British classical tradition (see \cite{Leijonhufvud1998} for an exact definition of the boundaries separating the classical and the modern or neoclassical traditions). This paper follows Keen's footsteps, with a big caveat. On the one hand, we endorse an accusation repeatedly brought against modern (\textit{neoclassical}) economic theory (known also as \textit{efficient resource allocation theory}), that is to be empirically and logically flawed.\footnote{Neoclassical economics is axiomatic. While it requires internal coherence, so that theorems can be logically deduced from a set of assumptions, it abstracts from external coherence between theoretical statements and empirical evidence. Of course, this implies an important epistemological detachment from falsifiable sciences like physics.} How the restricted vision of modern economics came to dominate the profession till our days is a fascinating chapter of the history of science which can not be addressed here (see \cite{Leijonhufvud2004,Mirowski2002}). On the other hand, however, we would like to underline that economics has been always characterized by the contemporaneous existence of several competing research traditions, which now and then jumble together. During the last sixty years it has not been unusual, for instance, to register that harsh criticisms to current mainstream conceptions were advanced by scholars who actively contributed to put them at the frontier of the economic discourse. In the \textit{war of the models} \cite{Freeman1998} fought by economists in the academic arena, neoclassicals have won several battles, but their opponents (\textit{e.g.}, Austrians, post-Keynesians, behavioralists) have never been completely defeated. On the contrary, from time to time these latter succeed in converting to their side some generals of the neoclassical army, who actively train recruits to innovative fighting techniques.
\par
The battleground we choose to exemplify our point is the research program launched some forty years ago by the neoclassical school, according to which macroeconomics should be explicitly grounded on microfoundations. Briefly, economic phenomena at a macroscopic level should be explained as the consequences of the activities undertaken by individual decision makers. The methodological strategy that has so far gained supremacy is one based on: \textit{i}) the precepts of the modern (\textit{i.e.}, rational choice-theoretic) tradition, and \textit{ii}) a solution (or \textit{closing}) concept borrowed from Walrasian competitive general equilibrium analysis.
\par
Some of the key flaws underlying item \textit{i}) has been presented in \cite{Keen2003}. In turn, critical words against item \textit{ii}) has been loudly pronounced, among the others, by eminent scholars who spent most of their academic life in the neoclassical camp \cite{HahnSolow1995}. Admittedly, their criticism was at one time well centered and largely neglected, probably because the alternative methodology they suggested was so close to the mainstream that the disturbing theoretical results they presented were seen as a nuisance, which could be easily addressed by painless manipulations of the standard model. The alternative proposal we shall confront with the Walrasian paradigm in what follows is definitely more radical. As a preliminary step, however, it seems worthwhile to review why Walrasian microfundations should be considered as the wrong answer to what is probably the most stimulating research question ever raised in economics, that is to explain how a completely decentralized economy composed of millions of (mainly) self-interested people coordinate their actions.


\section{Adam, L\'eon and the Butcher}
Contrary to what is usually thought, the very idea that the economy is a (complex) self-organizing system is not a new entry in the toolbox of economists got mixed up in complexity, but it is the key message conveyed in 1776 by the founding father of the discipline, the Scottish moral philosopher Adam Smith, according to whom:
\begin{quotation}
\emph{``He [man] generally, indeed, neither intends to promote the public interest, nor knows how much he is promoting it. By preferring the support of domestic to that of foreign industry, he intends only his own security; and by directing that industry in such a manner as its produce may be of the greatest value, he intends only his own gain, and he is in this, as in many other cases, led by an invisible hand to promote an end which was no part of his intention.}
\par
\emph{In civilized society he [man] stands at all times in need of the cooperation and assistance of great multitudes, while his whole life is scarce sufficient to gain the friendship of a few persons. In almost every other race of animals each individual, when it is grown up to maturity, is entirely independent, and in its natural state has occasion for the assistance of no other living creature. But man has almost constant occasion for the help of his brethren, and it is in vain for him to expect it from their benevolence only. He will be more likely to prevail if he can interest their self-love in his favour, and show them that it is for their own advantage to do for him what he requires of them. Whoever offers to another a bargain of any kind, proposes to do this. Give me that which I want, and you shall have this which you want, is the meaning of every offer; and it is in this manner that we obtain from one another the far greater part of those good offices which we stand in need of. It is not from the benevolence of the butcher, the brewer, or the baker that we expect our dinner, but from their regard to their self-love, and never talk to them of our own necessities but of their advantages.''} (Smith \cite{Smith1776}, p. 477 of the 1976 University of Chicago edition.)
\end{quotation}
\par
Five points deserve to be emphasized. First, the notion of an \textit{invisible hand} guiding coordination towards aggregate order in a fully decentralized economy represents an \textit{explanandum} (\textit{i.e.}, our research question) of theoretical and empirical importance, both for positive and normative reasons. Second, the explanation of economic phenomena is run in terms of the actions and reactions of autonomous individuals and not of social categories. This requires a conceptualization of decision-makers in terms of \textit{economic agents}, rather than social classes or races. Third, the economic agent (Smith's \textit{man}) envisaged in the quotation above responds to incentives to improve utility, but he is never requested to maximize it as he should do if endowed with substantive rationality. Fourth, Smith recognizes self-interest as being the main force driving decentralized actors towards a coordinated aggregate position, but he never concedes that greed is the only relevant human trait. Fifth, the work of the invisible hand yields a social order, but neither is such an order a rest (\textit{i.e.}, an equilibrium) nor is the best possible solution (\textit{i.e.}, an optimum).
\par
As regards the first two points\textemdash the key challenge of explaining how the invisible hand works and the approach of methodological individualism,\footnote{Actually, the path followed by methodological individualism has been rather bumpy, and what is nowadays commonly accepted by its very notion is somewhat different from the origins. After having brilliantly repelled the attack of Marxians, the currently established version of methodological individualism is illustrated by Arrow \cite{Arrow1994}, who acknowledges that individual behavior is always mediated by social relations, but also that social relations are the outcome of the actions of individual agents.} respectively\textemdash the profession has reached a consensus to which we convincingly adhere. Moving to the last three points, however, we subscribe to the far less popular claiming that the theoretical approach developed by the mainstream in dealing with them has proven to be inadequate. Instead of the idea of a self-regulating order emerging from the interactions of many simpler components without this \textit{``being part of their intentions''}, the Holy Trinity of (marginalist) neoclassical economics \cite{Colander2005}\textemdash (substantive) rationality, greed and equilibrium\textemdash has soon became the elected guiding principles of human behavior. In setting the methodological stage for the revolution\textemdash quickly turned into a dictatorship\textemdash dubbed Dynamic Stochastic General Equilibrium (DSGE) macroeconomic theory, Robert Lucas and Thomas Sargent bluntly declared:
\begin{quotation}
\emph{``An economy following a multivariate stochastic process is now routinely described as being in equilibrium, by which is meant nothing more that at each point in time (a) markets clears and (b) agents act in their own self-interest. This development, which stemmed mainly from the work of K. J. Arrow [...] and G. Debreu [...], implies that simply to look at any economic time series and conclude that it is a disequilibrium phenomenon is a meaningless observation. [...] The key elements of these models are that agents are rational, reacting to policy changes in a way which is in their best interests privately, and that the impulses which trigger business fluctuations are mainly unanticipated shocks.''} (Lucas and Sargent \cite{LucasSargent1979}, p. 7.)
\end{quotation}
\par
The self-regulating order of Smith was therefore transformed into a competitive General Equilibrium (GE) in the form elaborated in the 1870s by L\'eon Walras \cite{Walras1874}, that is a configuration of (fully flexible) prices and plans of action such that, at those prices, all agents can carry out their chosen plans and, consequently, markets clear. In a continuous effort of generalization and analytical sophistication, modern (neoclassical) economists interested in building microfundations for macroeconomics soon recurred to the refinement proposed in the 1950s by Arrow and Debreu \cite{ArrowDebreu1954}, who showed that also individual \textit{intertemporal} (on an infinite horizon) optimization yields a GE, as soon as the economy is equipped with perfect price foresights for each future state of nature and a complete set of Arrow-securities markets \cite{Arrow1964}, all open at time zero and closed simultaneously. Whenever these conditions hold true, the GE is an allocation that maximizes a properly defined social welfare function or, in other terms, the equilibrium is Pareto-efficient (First Welfare Theorem).\footnote{See \cite{IngraoIsrael1990} for a remarkable account of the origin and evolution of the GE concept.}
\par
As already anticipated, the weaknesses of the epistemological status of the GE model are so deep to have been fully recognized at various stages by its very proponents. Since this awareness still encounters huge difficulties in being spread among the profession, however, it seems worthwhile to provide a concise exposition of the main issues at hand.
\begin{enumerate}
\item The GE is neither unique nor locally stable under general conditions. This negative result, which refers to the work of Sonnenschein \cite{Sonneinschein1972}, Debreu \cite{Debreu1974} and Mantel \cite{Mantel1974}, can be summarized along the following lines. Let the aggregate excess demand function $F\left(p\right)$\textemdash obtained from aggregating among individual excess demands $f\left(p\right)$\textemdash be a mapping from the price simplex $\Pi$ to the commodity space $P^{N}$. A GE is defined as a price vector $p^{\ast}$ such that $F\left(p^{\ast}\right)=0$. It turns out that the only conditions that $F\left(\cdot\right)$ inherits from $f\left(\cdot\right)$ are continuity, homogeneity of degree zero and the Walras' law (\textit{i.e.}, the total value of excess demand is zero). These assure the existence, but neither the uniqueness nor the local stability of $p^{\ast}$, unless preferences generating individual demand functions are restricted to very special cases.
\item The existence of a GE is proved \textit{via} the Brower's fix point theorem, \textit{i.e.} by finding a continuous function $g\left(\cdot\right):\Pi\rightarrow\Pi$ such that any fix point for $g\left(\cdot\right)$ is also an equilibrium price vector $F\left(p^{\ast}\right)=0$. Suppose that we are interested in finding an algorithm, which, starting from an arbitrary price vector $p$, chooses price sequences to check for $p^{\ast}$ and halt when it finds it. In other terms, to find the GE price vector $F\left(p^{\ast}\right)=0$ means that halting configurations are decidable. As this violates the undecidability of the halting problem for Turing Machines, from a recursion theoretic viewpoint the GE solution is uncomputable \cite{RichterWong1999,Velupillai2000}. Notice that the same problem applies, in spite of its name, to the class of \textit{Computable} GE models \cite{Velupillai2005}.
\item By construction, in a GE all transactions are undertaken at the same equilibrium price vector. Economic theory has worked out two mechanisms capable to reach this outcome. First, one can assume that buyers and sellers adjust costlessly their optimal supplies and demands to prices called out by a (explicit or implicit) fictitious auctioneer, who continues to do his job until he finds a price vector which clears all markets. Only then transactions take place (Walras' assumption). Alternatively, buyers and sellers sign provisional contracts and are allowed to freely (\textit{i.e.}, without any cost) re-contract until a price vector is found which makes individual plans fully compatible. Once again, transactions occur only after the equilibrium price vector has been established (Edgeworth's assumption). Regardless of the mechanism one adopts, the GE model is one in which the formation of prices precedes the process of exchange, instead of being the result of it, through a \textit{t\^atonnement} process occurring in a meta-time. Real markets work the other way round and operates in real time, so that the GE model cannot be considered a scientific explanation of real economic phenomena \cite{Arrow1959}.
\item It has been widely recognized since \cite{Debreu1959} that integrating money in the theory of value represented by the GE model is at best problematic. No economic agent can individually decide to monetize alone; monetary trade should be the equilibrium outcome of market interactions among optimizing agents. The use of money\textemdash that is, a common medium of exchange and a store of value\textemdash implies that one party to a transaction gives up something valuable (for instance, his endowment or production) for something inherently useless (a fiduciary token for which he has no immediate use) in the hope of advantageously re-trading it in the future. Given that in a GE model actual transactions take place only after a price vector coordinating all trading plans has been freely found, money can be consistently introduced into the picture only if the logical keystone of the absence of transaction costs is abandoned. By the same token, since credit makes sense only if agents can sign contracts in which one side promises future delivery of goods or services to the other one, in equilibrium markets for debt are meaningless, both information conditions and information processing requirements are not properly defined, and bankruptcy can be safely ignored. Finally, as the very notion of a GE implies that all transactions occur only when individual plans are mutually compatible, and this has to be true also in the labor market, the empirically observed phenomenon of involuntary unemployment and the microfoundation program put forth by Lucas and Sargent are logically inconsistent.
\item	The only role assigned to time in a GE model is that of dating commodities. Products, technologies and preferences are exogenously given and fixed from the outset. The convenient implication of banning out-of-equilibrium transactions is simply that of getting rid of any disturbing influence of intermediary modifications of endowments\textemdash and therefore of individual excess demands\textemdash on the final equilibrium outcome.
\end{enumerate}
\par
The introduction of non-Walrasian elements into the GE microfoundations program\textemdash such as fixed or sticky prices, imperfect competition and incomplete markets leading to temporary equilibrium models\textemdash yields interesting Keynesian features such as the breaking of the Say's law and scope for a monetary theory of production, a rationale for financial institutions and a more persuasive treatment of informational frictions. As argued in \cite{Vriend1994}, however, all these approaches preserve a Walrasian perspective in that models are invariably closed by a GE solution concept which, implicitly or (more often) not, implies the existence of a fictitious auctioneer who processes information, calculates equilibrium prices and quantities, and regulates transactions. As a result, if the Walrasian Auctioneer (WA) is removed the decentralized economy becomes dynamically incomplete, as we are not left with any mechanism determining how quantities and prices are set and how exchanges occur.
\par
In turn, the flaws of the solution adopted by mainstream macroeconomists to overcome the problems of uniqueness and stability of equilibrium on the one hand, and of analytically tractability on the other one\textemdash\textit{i.e.}, the usage of a \textit{Representative Agent} (RA) whose choices summarize those of the whole population of agents\textemdash, are so pervasive and well known that it seems worthless to discuss them here. Interested readers are referred to \cite{Kirman1992,Hartley1997,GallegatiPalestriniDelliGattiScalas2006}.


\section{A Constructive Approach to Macroeconomics}
The research methodology we endorse in trying to explain successes and failures of the invisible hand consists in discarding the Walrasian GE approach to the microfoundation program, as well as its RA shorthand version. Instead of asking to deductively \textit{prove} the existence of an equilibrium price vector $p^{\ast}$ such that $F\left(p^{\ast}\right)=0$, we aimed at explicitly \textit{constructing} it by means of an \textit{algorithm} or \textit{rule}. From an epistemological perspective, this implies a shift from the realm of classical to that of constructive theorizing \cite{Velupillai2002}.\footnote{Epstein (in \cite{JuddTesfatsion2006}) prefers to talk of a generative approach to scientific explanation, but the meaning is basically the same.} Clearly, the act of computationally constructing a (fully or not) coordinated state\textemdash instead of imposing it \textit{via} the WA\textemdash for a decentralized economic system requires a complete description of goal-directed economic agents and their interaction structure.
\par
Agent-based computational economics (ACE)\textemdash that is the use of computer simulations to study evolving complex systems composed of many autonomous interacting agents\textemdash represents an effective implementation of such a research agenda \cite{JuddTesfatsion2006}. ACE allows an explicit modeling of identifiable, goal-directed, adapting agents, situated in an explicit space and interacting locally in it. In complex adaptive systems local interactions involves the spontaneous formation of macroscopic structures which can not be directly deduced by looking at individual behaviors. The equilibrium of a system does not require any more that every single element is in equilibrium by itself, but rather that the statistical distributions describing aggregate phenomena are stable, \textit{i.e.} in \emph{``[...] a state of macroscopic equilibrium maintained by a large number of transitions in opposite directions''} (\cite{Feller1957}, p. 356). A consequence of the idea that macroscopic phenomena can emerge is that the strong reductionist vision retained by neoclassicals is basically wrong. Once again, these concepts should be familiar to economists, at least to those who pay attention to the history of economic thought, since they mirror the notion of \textit{spontaneous market order} or \textit{catallaxy} put forth by the Nobel Prize in economics Friedrich von Hayek. According to him, a clear definition of the laws of property, tort and contract is enough to regulate a set of trial and error exchange relationships, which succeeds in coordinating the plans of an interdependent network of individuals endowed with a multiplicity of competing ends. In contrast, Hayek argues that the notion of competitive GE is \emph{``unfortunate, since it presupposes that the facts have already all been discovered and competition, therefore, has ceased''} (\cite{Hayek1978}, p. 184).
\par
It must be emphasized, however, that the abandonment of the GE solution concept does not come at no cost. The optimality of a competitive GE associated with the First Welfare Theorem implies that welfare comparisons between alternative macroeconomic states can be meaningfully carried out. An operational method for performing aggregate welfare comparisons also in macro ACE models has been advanced by Tesfatsion (in \cite{JuddTesfatsion2006}). The idea consists in defining an ideal benchmark (for instance, the GDP in correspondence of continual market clearing) and comparing it to the simulation outcomes. We can say the system has reached a \textit{catallaxy} if the distance between the simulated time series for GDP and the ideal reference path differs by less than a tolerance level $\tau$.


\section[The C@S Project]{The C@S Project\protect\footnote{\MakeLowercase{\MakeUppercase{T}he acronym stands for ``\MakeUppercase{C}omplex \MakeUppercase{A}daptive \MakeUppercase{T}rivial \MakeUppercase{S}ystem''. \MakeUppercase{F}or an exposition of the key issues and earlier results of this research program see [32].}}}
\noindent In \cite{GaffeoCatalanoDelliGattiGallegatiRusso2006} we \textit{grow} a sequential economy populated by a finite number of firms $i=1,\ldots,I$, workers/consumers $j=1,\ldots,J$, and banks $b=1,\ldots,B$, who undertake decisions at discrete times $t=1,\ldots,T$ on three markets, that is one for a homogeneous non-storable consumption good, one for labor services and one for credit services. Notional prices and quantities are chosen in an adaptive way, according to rules of thumb buffeted by idiosyncratic random disturbances. All three markets are characterized by decentralized search and matching processes, which imply individual, and \textit{a fortiori} aggregate, out-of-equilibrium dynamics. Thus, due to the absence of any exogenously imposed market-clearing mechanism, nothing prevents the economy from being characterized by a \textit{spontaneous order} and the contemporaneous occurrence of persistent involuntary unemployment, unsold production, excess individual demands and credit rationing.
\par
The sequence of events occurring in each period runs as follows.
\renewcommand{\theenumi}{\alph{enumi}}
\begin{enumerate}
\item At the beginning of any $t$, firms and banks check their financial viability as inherited from the past, and either they continue to operate if their neat wealth is positive, or if it is lower or equal to zero they shut down due to bankruptcy. In the latter case, a string of new firms/banks equal in number to the bankrupted ones enter the market. Entrants are simply random copies of incumbents.
\item Starting from the demand it expects to face, $D^{e}_{it}$, each operating firm determines the amount of output to be produced and the amount of labor to be hired. Expectations on future demand are updated adaptively:
\begin{equation}
D^{e}_{it}=\begin{cases}
D_{it-1}\left(1+\rho_{it}\right)&\text{if $D_{it-1}>Y_{it-1}$}\\
D_{it-1}&\text{if $D_{it-1}=Y_{it-1}$}\\
D_{it-1}\left(1-\rho_{it}\right)&\text{if $D_{it-1}<Y_{it-1}$}
\end{cases}
\end{equation}
where $D_{it-1}$ was the demand faced by firm $i$ during $t-1$, $Y_{it-1}$ was its supply, and $\rho_{it}$ is an idiosyncratic shock distributed on a positive support. The $i$\textsuperscript{th} firm sets $Y_{it}=D^{e}_{it}$ and, \textit{via} the CRS technology $Y_{it}=a_{it}L_{it}$, calculates its labor demand. Regardless of the scale of output, production takes one whole period.
\item A fully decentralized labor market opens. Firms set their wage bids (actually each of them sets a \textit{maximum wage}, as a function of its own financial soundness), and post their vacancies on the basis of their labor demand. Workers, in turn, accept a job only if the wage they are offered is higher than their individual satisfying wage. A sequential matching procedure determines whether unfilled vacancies and unemployed workers remain after the labor market has closed. Firms then pay their wage bill $W_{it}$ in order to start production.
\item If internal financial resources are in short supply for paying wages, firms can enter a fully decentralized credit market and fill in a fixed number of applications to obtain credit. We assume that due to prudential regulation, the total amount of loans supplied by each bank is proportional to its equity. Banks allocate credit collecting individual demands, sorting them in descending order according to the financial viability of firms, and satisfy them until all credit supply has been exhausted. The contractual interest rate is calculated applying a mark-up (function of financial viability) on a exogenously determined baseline.
\item After production is completed, a market for goods opens. Any of the $i$\textsuperscript{th} firm posts its offer price according to the adaptive rule:
\begin{equation}
P^{s}_{it}=\begin{cases}
P_{it-1}\left(1+\eta_{it}\right)&\text{if $D_{it-1}>Y_{it-1}$}\\
P_{it-1}&\text{if $D_{it-1}=Y_{it-1}$}\\
P_{it-1}\left(1-\eta_{it}\right)&\text{if $D_{it-1}<Y_{it-1}$}
\end{cases}
\end{equation}
where $\eta_{it}$ is an idiosyncratic random disturbance distributed on a positive support. Prices are not allowed to be lower than expected average costs, however. The demand for consumer goods is given by the whole labor income. Consumers muddle through the market searching for a satisfying deal. When the consumer $j$ identifies a firm with the lowest price in the neighborhood he can visit at no cost, he stops and concludes a transactions. If a firm ends up with excess supply, it gets rid of the unsold goods at zero costs.
\item Firms collect revenues, calculate profits, update their net worth and, if internal resources are enough, pay back their debt obligations. Firms employ part of retained profits in an R\&D investment, aimed at increasing the productivity parameter $a_{it}$. Returns from the R\&D investment are governed by a stochastic process.
\end{enumerate}
\par
The model is then completed with operational choices regarding procurement and matching processes, rationing methods, search costs, and statistical distributions for random disturbances. For instance, in structuring labor and goods markets we follow the lesson delivered by two distinguished  scholars, Arthur Okun \cite{Okun1981} and John Hicks \cite{Hicks1989}, who suggested to explicitly relate macroeconomic performance with non-Walrasian features like long-term employer-worker relationships and buyer-seller fidelity in customer markets. The great flexibility of the agent-based methodology we employ allows us to conduct experiments with alternative assumptions in a computable laboratory. We refer to \cite{GaffeoCatalanoDelliGattiGallegatiRusso2006} for all other technical details.
\par
\begin{figure}[!t]
\centering
\begin{center}
\mbox{
\subfigure[]{\label{fig:Figure1_a}\includegraphics[width=0.48\textwidth]{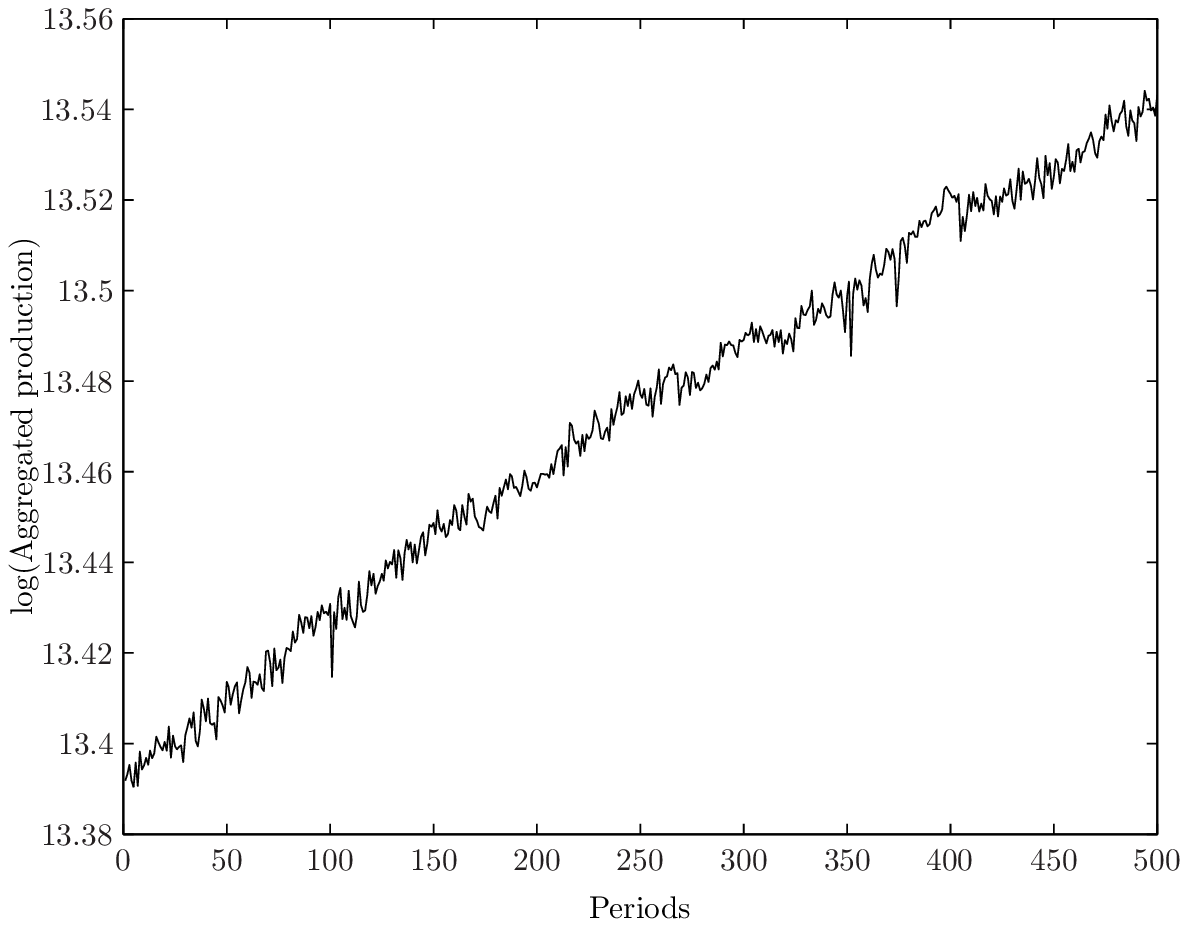}}
\subfigure[]{\label{fig:Figure1_b}\includegraphics[width=0.48\textwidth]{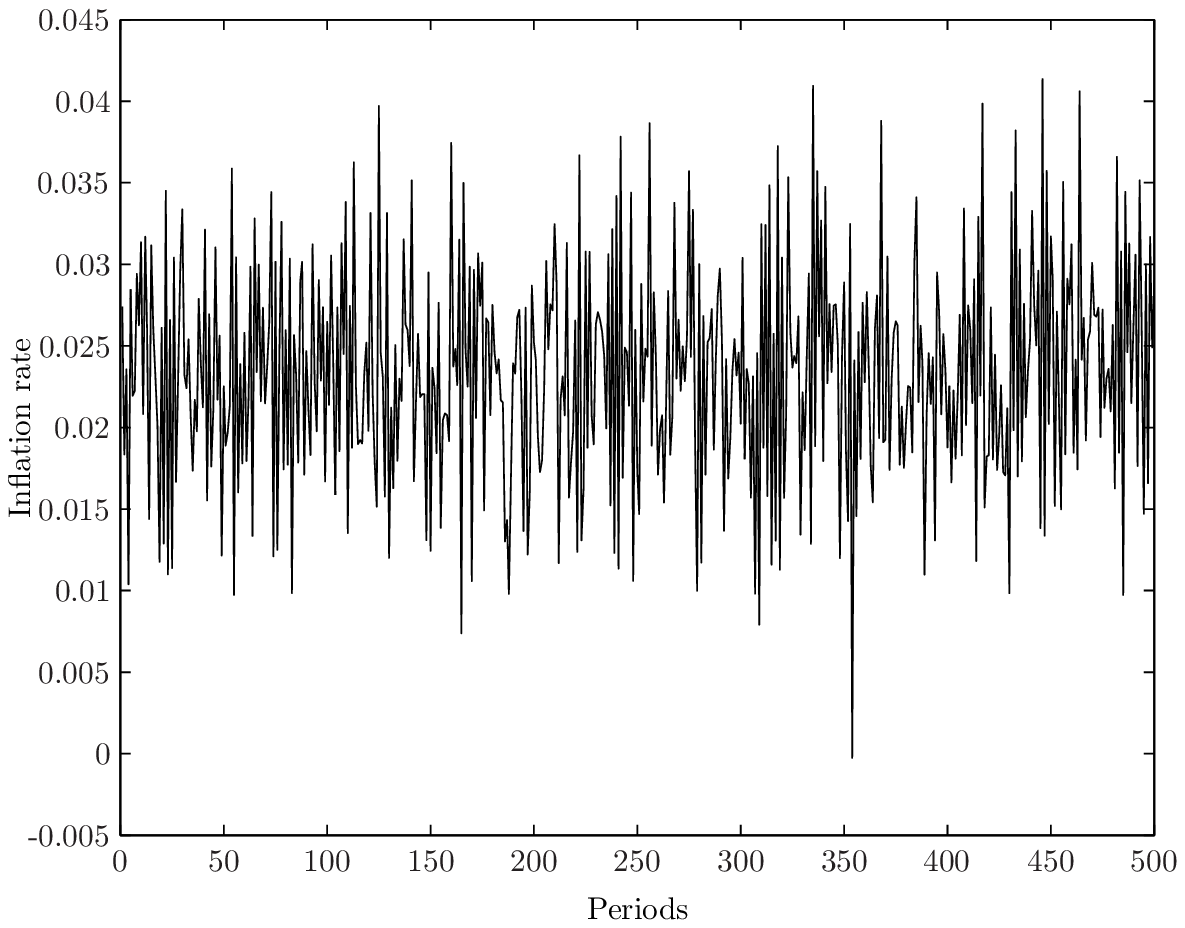}}
}
\mbox{
\subfigure[]{\label{fig:Figure1_c}\includegraphics[width=0.48\textwidth]{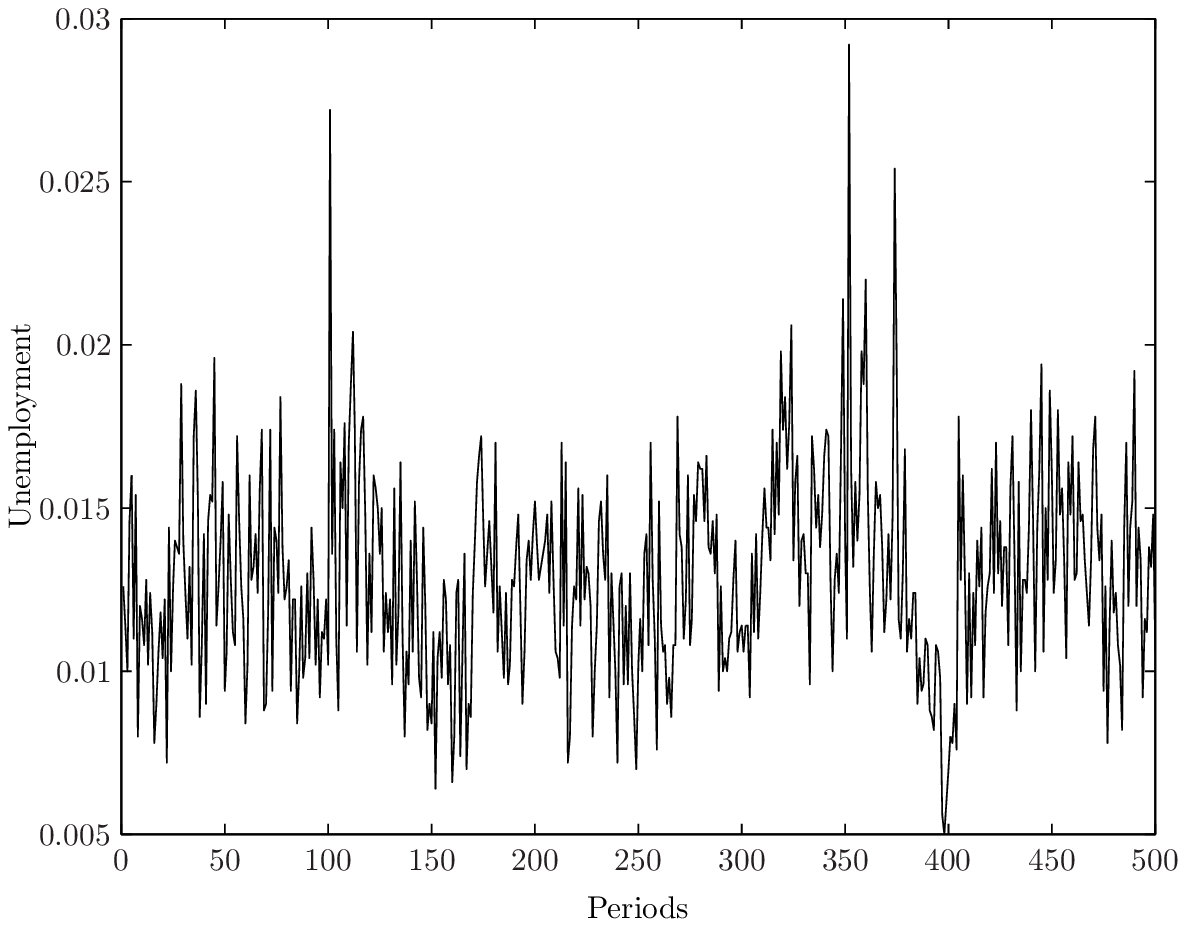}}
\subfigure[]{\label{fig:Figure1_d}\includegraphics[width=0.48\textwidth]{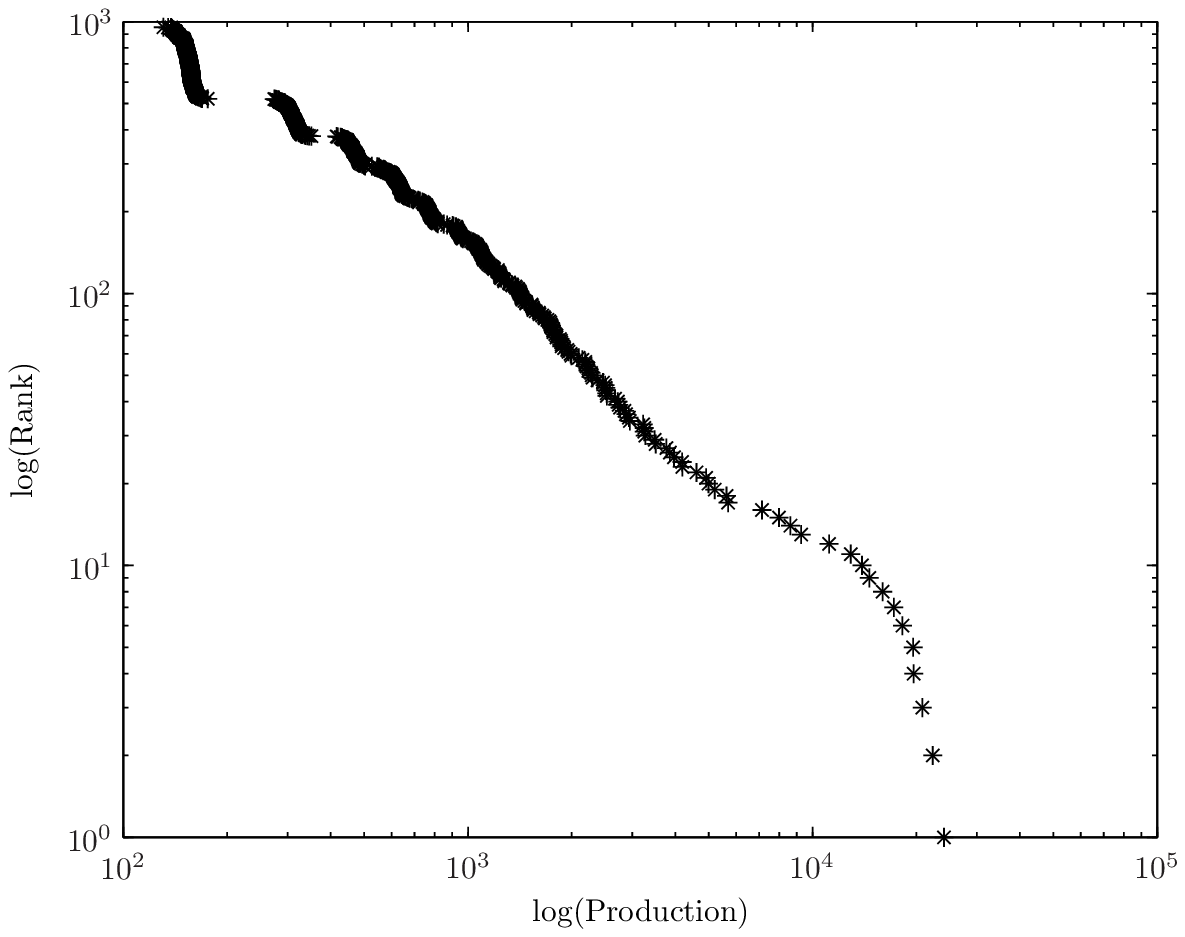}}
}
\caption{Macroeconomic regularities emerging from simulations: \subref{fig:Figure1_a} GDP (in log); \subref{fig:Figure1_b} inflation rate; \subref{fig:Figure1_c} unemployment rate; \subref{fig:Figure1_d} firms' size distribution}
\label{fig:Figure1}
\end{center}
\end{figure}

Panels \subref{fig:Figure1_a} to \subref{fig:Figure1_d} of FIG. \ref{fig:Figure1} present some selected results from simulations.
In spite of the absence of any imposed GE closing of the model, the aggregate dynamics emerging from decentralized market interactions among 1000 firms, 5000 worker/consumers and 100 banks is far from being violently instable. Aggregate output or GDP (Panel a) follows a sustained increasing path, and it mimics some well-known stylized facts of business cycle fluctuations, like alternating phases of smooth and large output variability, rare but significant crises (\textit{e.g.}, simulation periods 108--109, 354--362, 376--377), and segmented trends (\textit{e.g.}, periods 351--401 \textit{vs.} 402--500). Both the CPI inflation rate (Panel b) and the unemployment rate (Panel c) remain inside a stationary corridor. It seems interesting to note that the aggregate stability is fully consistent with wide heterogeneity. In Panel \subref{fig:Figure1_d}, for instance, we report the firms' size distribution as emerging at period 500 from an original uniform distribution. It clearly follows a power law, another renown regularity of industrial dynamics.


\section{Conclusion}
The preceding sections of this paper have been dedicated to four issues: 1) to introduce a key research question fascinating economists since the beginning of the discipline, that is that of explaining how millions of autonomous individuals with competing ends succeed in coordinating their economic activities by trading on fully decentralized markets; 2) to provide a critical assessment of the methodological solution endorsed by mainstream macroeconomists, who try to reconcile individual behaviors and aggregate economic phenomena (\textit{e.g.}, GDP, CPI, unemployment, income distribution) by means of Walrasian GE analysis; 3) to briefly discuss an alternative methodology, that is one rooted on an agent-based constructive approach; 4) to present some results from recent work undertaken along this latter and, according to us, much safer road to attain a reconciliation of micro and macroeconomic theory.
\par
In the remainder of this section, we shall discuss possible directions for future research. Three lines seem particularly promising. The first one consists in using our ACE macroeconomy as a computational laboratory in which market processes and institutions are allowed to grow endogenously from scratch. This requires an explicit treatment of transaction and communication costs, and a testable theory of how intermediaries (brokers and dealers) emerge. While interesting work in the ACE literature already exists (see \textit{e.g.} \cite{HowittClower2000,KirmanVriend2001}), our main concern is in assessing the endogenous co-evolution of markets, institutional arrangements and macroeconomic dynamics. For instance, an interesting question is how job creation and destruction and wage and price setting decisions at the firm level interact over the business cycle in correspondence of alternative market structures. 
The second line of future investigation will deal with validation exercises \cite{FagioloMonetaWindrum2006}. From this viewpoint, we can distinguish among: 1) \textit{descriptive output validation}, when simulated data are compared with data from real economies (\textit{e.g.}, macroeconomic time series and income distribution data for industrialized countries); 2) \textit{predictive output validation}, in which computationally generated output is to be matched with yet-to-acquired real data. Clearly, predictive validation must come with careful thinking about how to collect new data on possibly unexplored regularities; 3) \textit{input validation}, insuring that market micro-structures, institutional arrangements and behavioral rules incorporated into the model capture salient aspects of real-world economies. This requires an interplay of field-level data collection, human-based experiments and computational design \cite{Sunder2006}.
\par
Finally, the baseline model in its validated version can be used as a computational laboratory to compare the welfare-increasing performance of alternative monetary, fiscal or regulatory policies \cite{DelliGattiGaffeoGallegatiPalestrini2005,RussoCatalanoGaffeoGallegatiNapoletano2006}.



\begin{thebibliography}{breitestes Label}
\bibitem{Kirman2006}
Kirman A. P. (2006), Demand theory and general equilibrium: from explanation to introspection, a journey down the wrong road, mimeo, Princeton University.
\bibitem{Keen2003}
Keen S. (2003), Standing on the toes of pygmies: why econophysics must be careful of the economic foundations on which it builds, \textit{Physica A}, \textbf{324}:108--116.
\bibitem{Leijonhufvud1998}
Leijonhufvud A. (1998), Mr. Keynes and the Moderns, \textit{European Journal of the History of Economic Thought}, \textbf{1}:169--188.
\bibitem{Leijonhufvud2004}
Leijonhufvud A. (2004), The metamorphosis of neoclassical economics, in Bellet M., Gloria-Palermo S. and A. Zouache (eds.), \textit{Evolution of the Market Process: Austrian and Swedish Economics}. London, Routledge.
\bibitem{Mirowski2002}
Mirowski P. (2002), \textit{Machine Dreams. Economics Becomes a Cyborg Science}. Cambridge, Cambridge University Press.
\bibitem{Freeman1998}
Freeman R. (1998), War of the models: which labour market insititutions for the 21\textsuperscript{st} century?, \textit{Labour Economics}, \textbf{5}:1--24.
\bibitem{HahnSolow1995}
Hahn F. and R. Solow (1995), \textit{A Critical Essay on Modern Macroeconomic Theory}. Cambridge, MIT Press.
\bibitem{Smith1776}
Smith A. (1776), \textit{An Inquiry into the Nature and Causes of the Wealth of Nations}. New York, Modern Library.
\bibitem{Arrow1994}
Arrow K. J. (1994), Methodological individualism and social knowledge, \textit{American Economic Review}, \textbf{84}:1--9.
\bibitem{Colander2005}
Colander D. (2005), The future of economics: the appropriately educated in pursuit of the knowable, \textit{Cambridge Journal of Economics}, \textbf{29}:927--941.
\bibitem{LucasSargent1979}
Lucas R. and T. Sargent (1979), After Keynesian macroeconomics, \textit{Federal Reserve Bank of Minneapolis Quarterly Review}, \textbf{3}, Spring issue.
\bibitem{Walras1874}
Walras L. (1874), \textit{El\'ements d'\'economie politique pure}. Lausanne, Corbaz.
\bibitem{ArrowDebreu1954}
Arrow K. J. and G. Debreu (1954), Existence of an equilibrium for a competitive economy, \textit{Econometrica}, \textbf{22}:265--290.
\bibitem{Arrow1964}
Arrow K. J. (1964), The role of securities in the optimal allocation of risk-bearing, \textit{Review of Economic Studies}, \textbf{31}:91--96.
\bibitem{IngraoIsrael1990}
Ingrao B. and G. Israel (1990), \textit{The Invisible Hand. Economic Equilibrium in the History of Science}. Cambridge, MIT Press.
\bibitem{Sonneinschein1972}
Sonnenschein H. (1972), Market excess demand functions, \textit{Econometrica}, \textbf{40}:549--556.
\bibitem{Debreu1974}
Debreu G. (1974), Excess demand function, \textit{Journal of Mathematical Economics}, \textbf{1}:15--23.
\bibitem{Mantel1974}
Mantel R. (1974), On the characterization of aggregate excess demand, \textit{Journal of Economic Theory}, \textbf{7}:348--353.
\bibitem{RichterWong1999}
Richter M. K. and K. Wong (1999), Non-computability of competitive equilibrium, \textit{Economic Theory}, \textbf{14}:1--28.
\bibitem{Velupillai2000}
Velupillai K. V. (2000), \textit{Computable Economics}. Oxford, Oxford University Press.
\bibitem{Velupillai2005}
Velupillai K. V. (2005), The foundations of \textit{computable} general equilibrium theory, mimeo, University of Trento.
\bibitem{Arrow1959}
Arrow K. J. (1959), Towards a theory of price adjustment, in Abramovits M. (ed.), \textit{Allocation of Economic Resources}. Stanford, Stanford University Press.
\bibitem{Debreu1959}
Debreu G. (1959), \textit{The Theory of Value}. New York, John Wiley.
\bibitem{Vriend1994}
Vriend N. (1994), A new perspective on decentralized trade, \textit{Economie Appliqu\'ee}, \textbf{46}:5--22.
\bibitem{Kirman1992}
Kirman A. P. (1992), Whom or what does the representative individual represent?, \textit{Journal of Economic Perspectives}, \textbf{6}:117--136.
\bibitem{Hartley1997}
Hartley J. E. (1997), \textit{The Representative Agent in Macroeconomics}. London, Routledge.
\bibitem{GallegatiPalestriniDelliGattiScalas2006}
Gallegati M., Palestrini A., Delli Gatti D. and E. Scalas (2006), Aggregation of heterogeneous interacting agent: the variant representative agent framework, \textit{Journal of Economic Interaction and Coordination}, \textbf{1}:5--19.
\bibitem{Velupillai2002}
Velupillai K. V. (2002), Effectivity and constructivity in economic theory, \textit{Journal of Economic Behavior and Organization}, \textbf{49}:307--325.
\bibitem{JuddTesfatsion2006}
Judd K. L. and L. S. Tesfatsion (2006), \textit{Handbook of Computational Economics, Vol. 2: Agent-Based Computational Economics}. Amsterdam, North Holland.
\bibitem{Feller1957}
Feller W. (1957), \textit{Introduction to Probability Theory and its Applications, Vol. 1}. New York, Wiley.
\bibitem{Hayek1978}
Hayek F. A. (1978), \textit{New Studies in Philosophy, Politics, Economics and the History of Ideas}. Chicago, University of Chicago Press.
\bibitem{DelliGattiGaffeoGallegatiGiulioniPalestrini2006}
Delli Gatti D., Gaffeo E., Gallegati M., Giulioni G. and A. Palestrini (2006), \textit{Emergent Macroeconomics. An Agent-based Approach to Business Fluctuations}. Forthcoming.
\bibitem{GaffeoCatalanoDelliGattiGallegatiRusso2006}
Gaffeo E., Catalano M., Delli Gatti D., Gallegati M. and A. Russo (2006), Macroeconomic dynamics in an agent-based model, mimeo, Universit\`a Politecnica delle Marche.
\bibitem{Okun1981}
Okun A. (1981), \textit{Prices \& Quantities. A Macroeconomic Analysis}. Washington, The Brookings Institution.
\bibitem{Hicks1989}
Hicks J. (1989), \textit{A Market Theory of Money}. Oxford, Clarendon Press.
\bibitem{HowittClower2000}
Howitt P. and R. Clower (2000), The emergence of economic organization, \textit{Journal of Economic Behavior and Organization}, \textbf{41}:55--84.
\bibitem{KirmanVriend2001}
Kirman A. P. and N. Vriend (2001), Evolving market structure: an ACE model of price dispersion and loyalty, \textit{Journal of Economic Dynamics and Control}, \textbf{25}:459--502.
\bibitem{FagioloMonetaWindrum2006}
Fagiolo G., Moneta A. and P. Windrum (2006), Empirical validation of agent-based models, \textit{Journal of Economic Behavior and Organization}, in press.
\bibitem{Sunder2006}
Sunder S. (2006), Determinants of economic interaction: behavior and structure, \textit{Journal of Economic Interaction and Coordination}, \textbf{1}:21--32.
\bibitem{DelliGattiGaffeoGallegatiPalestrini2005}
Delli Gatti D., Gaffeo E., Gallegati M. and A. Palestrini (2005), The apprentice wizard: monetary policy, complexity and learning, \textit{New Mathematics and Natural Computation}, \textbf{1}:109--128.
\bibitem{RussoCatalanoGaffeoGallegatiNapoletano2006}
Russo A., Catalano M., Gaffeo E., Gallegati M. and M. Napoletano (2006), Industrial dynamics, fiscal policy and R\&D: evidence from a computational experiment, \textit{Journal of Economic Behavior and Organization}, in press.
\end{thebibliography}
\end{document}